\documentclass[12pt]{article}
\usepackage{amsmath,amssymb,amscd}
\usepackage{epsfig}
\begin{document}

\bigskip
\bigskip

\begin{center}
{\bf MARKET MILL DEPENDENCE PATTERN IN THE STOCK MARKET: INDIVIDUAL PORTRAITS}
\end{center}

\bigskip

\begin{center}
\bf{ \large Andrei Leonidov$^{(a,b,c)}$\footnote{Corresponding author. E-mail leonidov@lpi.ru}, Vladimir Trainin$^{(a)}$, \\Alexander
Zaitsev$^{(a)}$, Sergey Zaitsev$^{(a)}$}
\end{center}
\medskip

(a) {\it Letra Group, LLC, Boston, Massachusetts, USA}\\

(b) {\it Theoretical Physics Department, P.N.~Lebedev Physics Institute,\\
    Moscow, Russia}

(c) {\it Institute of Theoretical and Experimental Physics, Moscow, Russia}

\bigskip

\bigskip

\bigskip

\begin{center}
{\bf Abstract}
\end{center}

This paper continues a series of studies of dependence patterns following from properties of the bivariate
probability distribution ${\cal P}(x,y)$ of two consecutive price increments $x$ (push) and $y$ (response). The
paper focuses on individual differences of the ${\cal P}(x,y)$ for 2000 stocks using a methodology of
identification of asymmetric market mill patterns developed in \cite{LTZZ05a,LTZZ05b}. We show that individual
asymmetry patterns (portraits) are remarkably stable over time and can be classified into three major groups -
correlation, anticorrelation and market mill. We analyze the conditional dynamics resulting from the properties of
${\cal P}(x,y)$ for all groups and demonstrate that it is trend-following at small push magnitudes and contrarian
at large ones.

\newpage

\section{Introduction}

Development of a quantitative picture of the dynamics of financial markets essentially consists in unraveling
dependence patterns characterizing the evolution of prices. Inherent randomness present in price dynamics leads to
a necessity of using the probabilistic approach. The full probabilistic description of dependence patterns is then
provided by the corresponding multivariate probability distributions. In the simplest case considered in the
present paper and the previous papers of this series \cite{LTZZ05a,LTZZ05b} we analyze the dependence patterns
linking two consecutive price increments.

An idea of describing the stock price dynamics in terms of the bivariate probability density ${\cal P}(x,y)$
characterizing the probabilistic interrelation between two consecutive price increments, $x$ (push) and $y$
(response), was first explicitly suggested by Mandelbrodt \cite{M63}, where interesting features of bivariate Levy
distribution were considered. An idea of characterizing individual dynamical properties of a single stock by the
two-dimensional portrait of the equiprobability lines of this distribution was later put forward in \cite{MB}. The
properties of the bivariate probability distribution ${\cal P}(x,y)$  were also discussed in relation to the
so-called "compass rose" pattern \cite{CL96}, in particular, in connection with the issue of predictability
\cite{AV04,V04}.

The present paper continues a series of papers \cite{LTZZ05a,LTZZ05b} devoted to an analysis of dependence
patterns following from particular properties of the bivariate distribution ${\cal P}(x,y)$. In the previous
papers \cite{LTZZ05a,LTZZ05b} we studied an average picture describing properties of an ensemble of pairs of
consequent price increments combining all stocks under consideration. Detailed studies of high frequency data
revealed a complex nonlinear probabilistic dependence pattern linking consecutive price increments $x$ and $y$. In
the first paper of the series \cite{LTZZ05a} we discussed various asymmetries characterizing the bivariate
distribution and showed that all of them are described by the same universal market mill pattern. In the next
paper \cite{LTZZ05b} we performed a detailed analysis of the geometry of the bivariate distribution characterizing
the consecutive price increments. We found, in particular, that shapes of the bivariate distribution around the
origin in the $xy$ plane and far from it are noticeably different. In particular, the conditional response
distribution ${\cal P}(y|x)$ at increasing push magnitude was found to become progressively more gaussian.

In the present paper we analyze the dependence patterns characterizing the bivariate distribution ${\cal P}(x,y)$
on the stock-by-stock basis.

The paper is organized as follows.

Section~2 is devoted to an identification of the basic types of asymmetry patterns of the bivariate distribution
for individual stocks, analysis of the frequency of their occurrence and of the characteristic features of
conditional dynamics corresponding to different pattern types. In paragraph 2.1 we describe the dataset and the
probabilistic methodology used in our analysis. In paragraph 2.2.1 we analyze the basic features of the summary
bivariate probability distribution studied in the previous papers \cite{LTZZ05a,LTZZ05b}. In paragraph 2.2.2 we
introduce several visually recognizable basic types of bivariate dependence patterns characterizing individual
stocks: correlation, anticorrelation and market mill ones. In paragraph 2.2.3 a quantitative classification of
individual bivariate patterns based on the probabilistic weight of various sectors in $xy$ plane computed on the
annual basis is introduced. An analysis of the ensembles of stocks traded at NYSE and NASDAQ shows important
differences in the relative proportions of various patterns. In paragraph 2.2.4 we show that different patterns
are in fact "deformed" variants of the basic market mill pattern and have common basic features of conditional
dynamics which is trend-following at small pushes and contrarian at large ones.  In paragraph 2.2.5 we analyze the
relative yield of different patterns on the monthly basis. We show that for the overwhelming majority of stocks
conditional dynamics is characterized by two coexisting types of patterns that are on average mixed at a ratio
3:1. In paragraph 2.2.6 we discuss the asymmetry patterns characterizing non-tradable and tradable indices. We
show that the non-tradable indices are characterized by the correlation dependence pattern while the tradable
indices are characterized by the anticorrelation one.

In section~3 we turn to an analysis of the pattern stability on the monthly time scale.  We introduce a distance
in the pattern space and show that the average distance between the consecutive monthly patterns of the same stock
is less than that between the considered stock and all other stocks or all other stocks in the same pattern group
in the same month.

In section~4 we formulate the main conclusions of the paper.

\section{Individual portraits}

\subsection{Data and methodology}

Our analysis is based on the data on 1 -minute price increments of 2000 stocks traded at NYSE and NASDAQ stock exchanges in the year of
2004\footnote{All holidays, weekends and off-market hours are excluded from the dataset.}.

We consider two consequent time intervals of length $\Delta T$, correspondingly $x=p(t_1 + \Delta T)-p(t_1)$ (push) and $y=p(t_1 + 2 \, \Delta
T)-p(t_1+\Delta T)$ (response). The full probabilistic description of price dynamics in the push-response $xy$ plane is given by a bivariate
probability density ${\cal P} (x,y)$. Knowledge of the bivariate distirbution ${\cal P} (x,y)$ fully specifies, in particular, a description of the
corresponding conditional dynamics in terms of the conditional distribution ${\cal P}(y |\, x) = {\cal P}(x,y)/{\cal P}(x)$.

In the previous papers \cite{LTZZ05a,LTZZ05b} we studied the properties of the summary bivariate probability distribution ${\cal P}_{\rm tot} (x,y)$
for the complete ensemble of pairs of price increments combining contributions from all the stocks considered.

In the present paper we focus on analyzing the properties of the bivariate probability distributions ${\cal P}_i(x,y)$, where $i$ indexes some
particular stock.

\subsection{Individual portraits: identification and statistics}

As stated above, our goal is in studying the properties of bivariate probability distributions ${\cal P}_i(x,y)$ that fully characterizes the
probabilistic pattern linking consecutive price increments $x$ and $y$ for the $i$-th stock. We shall refer to these patterns as to "individual
portraits".

\subsubsection{Group portrait}

Let us start with reminding of how the "group portrait", i.e. the bivariate distribution ${\cal P}_{\rm tot}(x,y)$ combining all pairs of consecutive
price increments considered, looks like \cite{LTZZ05a,LTZZ05b}. A two-dimensional projection of $\log_8 {\cal P}_{\rm tot}(x,y)$ for the particular
case of consecutive 1 - minute price increments is shown in Fig.~1.
\begin{figure}[h]
 \begin{center}
 \leavevmode
 \epsfysize=11cm
 \epsfbox{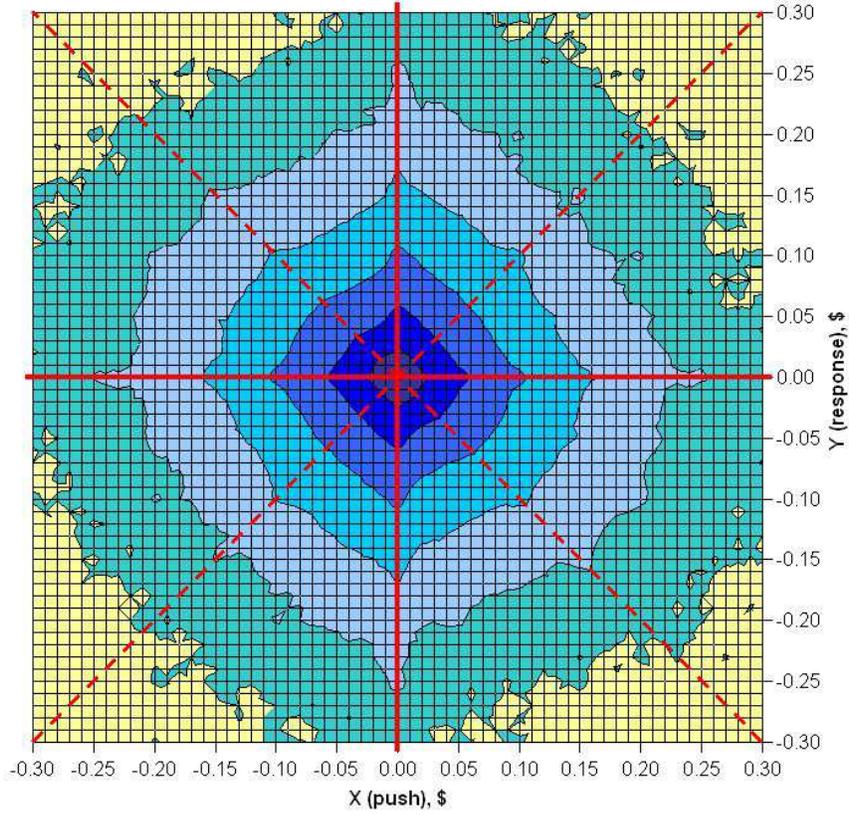}
 \end{center}
\caption{Logarithm of two-dimensional distribution $\log_8 {\cal P}_{\rm tot}(x,y)$, $\Delta T=1$ min. }
\end{figure}
As shown in \cite{LTZZ05a,LTZZ05b}, this distribution has a number of remarkable properties illustrated by the sketch of the profile of
equiprobability levels in Fig.~2,
\begin{figure}[h]
 \begin{center}
 \leavevmode
 \epsfysize=7cm
 \epsfbox{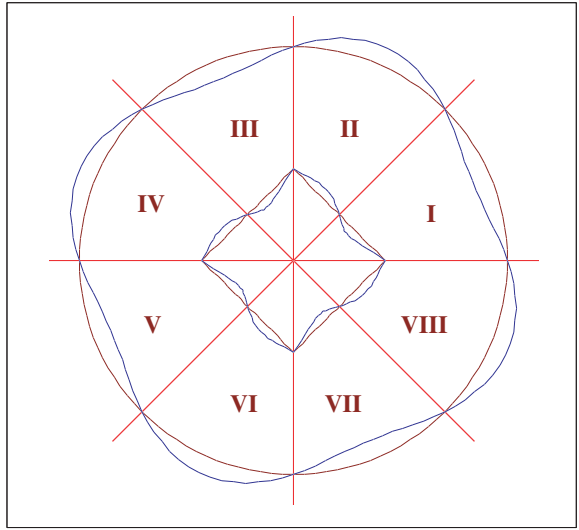}
 \end{center}
\caption{Sketch of the equiprobability levels of the bivariate distribution ${\cal P}_{\rm tot}(x,y)$. The basic regular symmetric structure is shown
by brown lines, the actual structure - by blue ones. }
\end{figure}
in which the push-response $xy$ plane is divided into sectors numbered counterclockwise from I to VIII.

A detailed description of symmetries of the bivariate probability distribution ${\cal P}_{\rm tot}(x,y)$ is achieved \cite{LTZZ05a} by considering
its symmetry properties with respect to the axes $y=0$, $y=x$, $x=0$ and $y=-x$ . In the present paper we shall restrict our consideration to the
asymmetry with respect to the axis $x=0$\footnote{An analysis of the asymmetry pattern with respect to the axis $y=x$ shows that the arising
individual portraits are either of the market mill type or noise.}.

A convenient visualization of the symmetry pattern in question is provided by the following decomposition \cite{LTZZ05a,LTZZ05b} of ${\cal P}_{\rm
tot} (x,y)$ into symmetric and antisymmetric components with respect to the axis  $y=0$:
\begin{equation}\label{sepdis1}
 {\cal P}_{\rm tot} (x,y) \, = \, {\cal P}^s_{\rm tot} (x,y)+ {\cal P}^a_{\rm tot} (x,y)
\end{equation}
where
\begin{equation}\label{sepdis11}
{\cal P}^s_{\rm tot} \, = \, \frac{1}{2} \left ( {\cal P}_{\rm tot} (x,y) + {\cal P}_{\rm tot}(x,-y)  \right ) \,\,\,\,\, {\rm and} \,\,\,\,\, {\cal
P}^a_{\rm tot} \, = \, \frac{1}{2} \left ( {\cal P}_{\rm tot} (x,y) - {\cal P}_{\rm tot}(x,-y)  \right )
\end{equation}
In principle both symmetric ${\cal P}^s_{\rm tot}$ and antisymmetric ${\cal P}^a_{\rm tot}$ components can be used for identifying the basic types of
individual portraits. In the present paper we shall restrict ourselves to considering the antisymmetric component ${\cal P}^a_i$ only. To visualize
the asymmetric contribution it convenient to consider \cite{LTZZ05a} its positive part
\begin{equation}\label{sepdis12}
{\cal P}_{\rm tot}^{a (p)}(x,y) = \theta \left[ {\cal P}_{\rm tot} ^{a}(x,y) \right] \cdot {\cal P}_{\rm tot}^{a}(x,y),
\end{equation}
 where $\theta [ \cdot ]$ is a
Heaviside step function. The definition of ${\cal P}_{\rm tot} ^{a (p)}(x,y)$ guarantees that no information is lost when imposing this restriction.

A salient feature of the distribution shown in Fig.~1 and sketched in Fig.~2 is that all even sectors are stronger (contain more of probability
density) than all odd ones. This leads to the remarkable market mill structure of the summary distribution ${\cal P}_{\rm tot}^{a (p)}(x,y)$ shown in
Fig.~3
\begin{figure}[h]
 \begin{center}
 \leavevmode
 \epsfysize=11cm
 \epsfbox{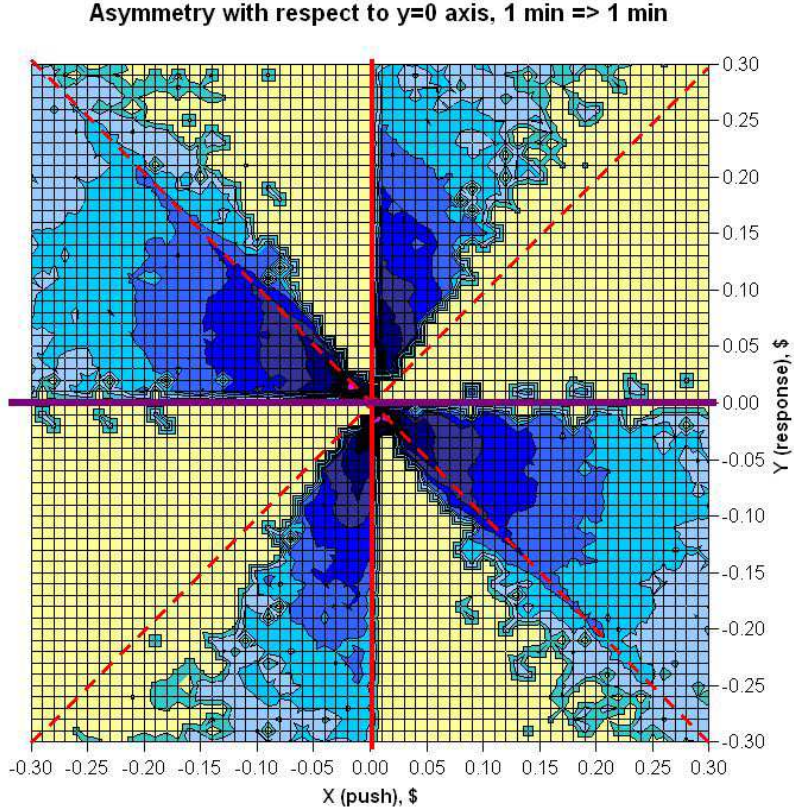}
 \end{center}
\caption{Asymmetry of two-dimensional distribution $\log_8 {\cal P}_{\rm tot}^{a (p)}(x,y)$, $\Delta T=1$ min. }
\end{figure}
which leads, in particular, to the $z$ - shaped mean conditional response $\langle y \rangle_x$ described in \cite{LTZZ05a}. Let mention two
important features characterizing the distribution plotted in Fig.~3. First, the blades in the second and fourth quadrants corresponding to negative
autocorelation are more pronounced than their correlation counterparts in the first and third quadrants of the $xy$ plane. A second interesting
feature is a a presence of a number of isolated small domains, in particular right below the negative and right above the positive semi-axes $x<0$
and $x>0$. Their existence can be explained by the specific asymmetry related to the rounding of the summary increment $x+y$ with \$ 0.05 accuracy,
so that, e.g. a sequence of increments \$ 0.08, \$ -0.18 will be more frequent than \$ 0.08, \$ 0.18. Let us note that had we worked with normalized
increments, i.e. returns, we would not be able to see this effect.

\subsubsection{Basic patterns}\label{visclas}

Is the pattern shown in Figs.~1,2,3 a dominant one for all the stocks or does it result from a superposition of several distinctly different types of
portraits? To answer this question we have to try to identify a robust basis in the pattern space and analyze the frequencies with which these basic
patterns occur. In what follows we shall concentrate on analyzing the asymmetry of individual bivariate probability distributions ${\cal P}_i (x,y)$,
$i=1, \cdots, 2000$, as characterized by the positive part of its asymmetric component ${\cal P}_i^{a (p)}(x,y)$ defined analogously to the case of
summary distribution, see Eqs.~(\ref{sepdis1}) - (\ref{sepdis12}).

The market mill shape of the group distribution asymmetry shown in Fig.~3 looks especially nontrivial if compared to standard asymmetry patterns
corresponding to correlated (anticorrelated) push $x$ and response $y$, which just correspond to symmetric patterns filling quadrants 1 and 3 and 2
and 4 respectively. It is tempting to interpret the appearance of the market mill pattern as resulting from  a blend of correlative and
anticorrelative behaviors. A useful starting point for the subsequent pattern type analysis can thus be a separation of the pattern space into three
groups corresponding to the dominance of market mill, correlation and anticorrelation behavior in the conditional dynamics of an individual stock.

Visual inspection of the geometric structure of the individual portraits ${\cal P}_i^{a (p)}(x,y)$ indeed suggests that a vast majority of them can
be classified as belonging to these three main types. The sample patterns for the stocks DIS, HDI and DE providing clear examples of anticorrelation
(DIS), market mill (HDI) and correlation (DE) behavior are shown in Fig.~4.
\begin{figure}[h]
 \begin{center}
 \leavevmode
 \epsfysize=4.7cm
 \epsfbox{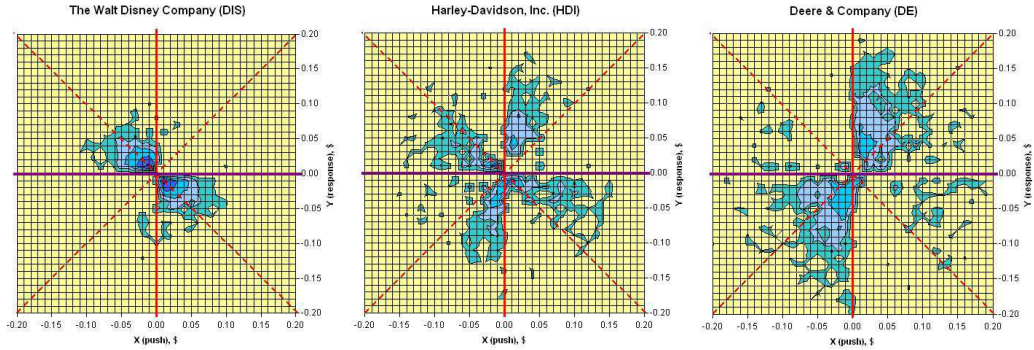}
 \end{center}
\caption{Asymmetry portraits for DIS, HDI and DE stocks.}
\end{figure}

\subsubsection{Basic patterns: quantitative classification}

Visual pattern recognition, albeit being very effective, is not the most convenient way of classifying a large number of images. It is therefore
necessary to devise some simple quantitative measure providing a sufficiently robust separation of portraits into well-recognizable groups. To
construct such quantitative framework for distinguishing various pattern types let us return to the full bivariate distributions ${\cal P}_i (x,y)$
and consider a rectangular domain in the $xy$ plane $ \{ |x| \leq \$ \,\, 0.3, |y| \leq  \$ \,\, 0.3 \}$. Our classification will be based on the
relative weight of the eight sectors shown in  Fig.~2. To calculate these weights we first remove all points lying exactly on the borders between the
sectors and denote the relative weights of the points lying inside the sectors I -- VIII by $w_{\rm I}, \cdots , w_{\rm VIII}$ correspondingly. A
useful way of graphically presenting a pattern characterized by a particular set of weights $w_{\rm I}, \cdots , w_{\rm VIII}$ is a spider's web
diagram shown in Fig.~5 for the same three representative stocks DIS, HDI and DE (see Fig.~4).
\begin{figure}[h]
 \begin{center}
 \leavevmode
 \epsfysize=3.8cm
 \epsfbox{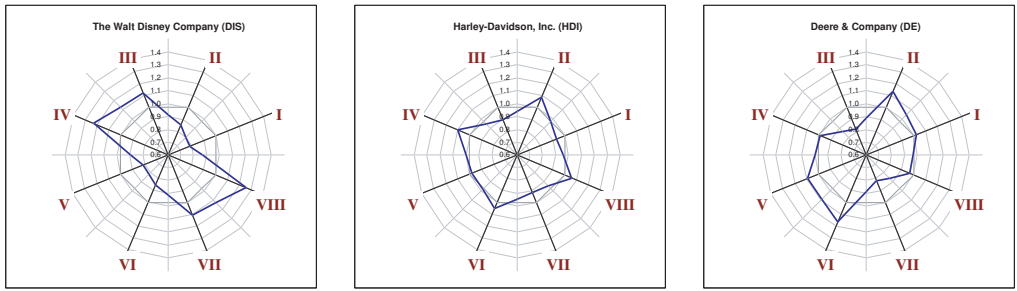}
 \end{center}
\caption{Spider's web patterns for DIS, HDE and DE stocks representing the anticorrelation, market mill and correlation patterns respectively.}
\end{figure}
In this diagram the weights are marked on the bisectors of the corresponding sectors. We see that different basic patterns correspond to distinctly
different spider web diagrams. The anticorrelation pattern (DIS) is located along the north-west -- south-east axis, the correlation one (DE) is
located along the north-east -- south-west axis, while the market mill one (HDE) can be described as a skewed rhomboid.

The group portrait pattern shown in Fig.~1 and sketched in Fig.~2 corresponds to a situation in which all even sectors are stronger than all odd
ones, so that on the set of sector weights $w_{\rm I}, \cdots , w_{\rm VIII}$ this generates a system of 16 inequalities. We have already stressed
that these are valid for a summary probability distribution including all stocks, while at an individual there exist several patterns characterized
by some particular ordering of the frequencies $w_{\rm I}, \cdots , w_{\rm VIII}$.  A detailed analysis shows that at an individual level the
above-mentioned 16 conditions weaken in such a way that
\begin{itemize}
 \item{out of 16 inequalities valid for the summary distribution (all even sectors are stronger than all odd ones) there remains only four
 \footnote{This is true for about 90$\%$ of all stocks. Note that visual inspection of Fig.~5 confirms the validity of inequalities (\ref{fourineq})
  for all three stocks presented in the figure.}:
 \begin{equation}\label{fourineq}
 w_{\rm II} > w_{\rm I} \,\,\,\,\ w_{\rm IV} > w_{\rm III} \,\,\,\,\, w_{\rm VI} > w_{\rm V} \,\,\,\,\ w_{\rm VIII} > w_{\rm VII}
 \end{equation}
 }
 \item{The distribution is centrally symmetric:
 \begin{equation}\label{foureq}
 w_{\rm I} = w_{\rm V} \,\,\,\, w_{\rm II} = w_{\rm VI} \,\,\,\, w_{\rm III} = w_{\rm VII} \,\,\,\, w_{\rm IV} = w_{\rm VIII}
 \end{equation}
 }
\end{itemize}
Note that because of the central symmetry of the distribution it is sufficient to consider a positive half-plane $x>0$. All possible configurations
are specified by ordered permutations of the sequence $\left( w_{\rm I}, w_{\rm II}, w_{\rm III},w_{\rm IV} \right)$ under conditions $w_{\rm II} >
w_{\rm I}$ and $w_{\rm VIII} > w_{\rm VII}$. Therefore the conditions (\ref{fourineq},\ref{foureq}) allow for six possible orderings of weights. To
project this information on the desired classification into three types (correlation, anticorrelation and market mill) we identify each pattern with
a point in $AC$ plane, where
\begin{eqnarray}
 A & = & \left( w_{\rm II} +w_{\rm VI} \right ) - \left( w_{\rm III} +w_{\rm VII} \right ) \nonumber \\
 C & = & \left( w_{\rm IV} +w_{\rm VIII} \right ) - \left( w_{\rm I} +w_{\rm V} \right )
\end{eqnarray}
It is easy to check that the three above-discussed patterns described in the previous paragraph can be identified with the following location in the
$AC$ plane:
\begin{itemize}
\item{Quadrant I \,\, ($A>0$, $C>0$). Market mill (MILL)}
\item{Quadrant II \, ($A<0$, $C>0$). Negative autocorrelation (ACOR)}
\item{Quadrant III \, ($A<0$, $C>0$). Anti-mill (AMILL)}\footnote{The "Anti-mill" pattern would visually look like the market mill one
rotated clockwise at $\pi/4$. In terms of sector weights this would correspond to even sectors being stronger than odd ones. Rather remarkably, such
pattern never appears.}
\item{Quadrant IV  ($A>0$, $C<0$). Positive autocorrelation (COR)}
\end{itemize}
In Fig.~6 we show the positions of the three sample stocks DIS, HDI and DE in the $AC$ plane. The shaded region shows the domain in which almost all
of points corresponding to 2000 stocks under consideration lie.
\begin{figure}[h]
 \begin{center}
 \leavevmode
 \epsfysize=10cm
 \epsfbox{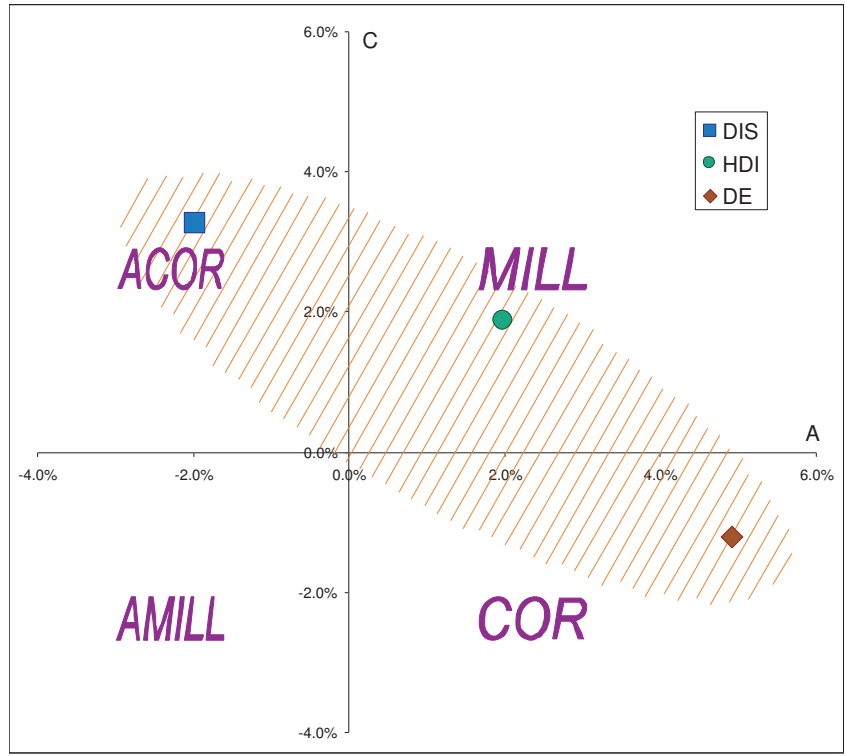}
 \end{center}
\caption{Positions of DIS, HDI and DE in the $AC$ plane. The shaded region shows the domain in which alsmost all $AC$ images lie.}
\end{figure}

In terms of weight orderings there emerges the following classification of six possible weight orderings into types:

\medskip

\begin{center}
{\bf Table 1}
\end{center}
\begin{center}
\begin{tabular}{|c|c|}
  \hline
  $w_{\rm II} > w_{\rm I} > w_{\rm VIII} > w_{\rm VII}$ & Correlation \\ \hline
  $w_{\rm II} > w_{\rm VIII} > w_{\rm I} > w_{\rm VII}$ & Mill \\ \hline
  $w_{\rm II} > w_{\rm VIII} > w_{\rm VII} > w_{\rm I}$ & Mill \\ \hline
  $w_{\rm VIII} > w_{\rm II} > w_{\rm I} > w_{\rm VII}$ & Mill \\ \hline
  $w_{\rm VIII} > w_{\rm II} > w_{\rm VII} > w_{\rm I}$ & Mill \\ \hline
  $w_{\rm VIII} > w_{\rm VII} > w_{\rm II} > w_{\rm I}$ & Anticorrelation \\ \hline
\end{tabular}
\end{center}
\begin{center}
Table~1. {\small Emergence of individual patterns in terms of weight orderings.}
\end{center}
the type (ACOR, MILL or COR) being assigned to each configuration by identifying the two strongest sectors. From classification of configurations
presented in Table~1 we conclude that provided the inequalities (\ref{fourineq}) and equalities (\ref{foureq}) take place, correlation and
anticorrelation patterns appear in 1/6 of cases each, while the dominant 2/3 correspond to the market mill pattern.

On the one-year horizon the total ensemble of 2000 stocks is characterized by the following pattern decomposition for the year 2004:

\medskip
\begin{center}
{\bf Table 2}
\end{center}
\begin{center}
\begin{tabular}{|c|c|c|c|c|}
\hline
    {\bf Type} & {\bf ACOR} & {\bf MILL} & {\bf COR} & {\bf AMILL} \\ \hline
    {\bf Number} & 777 & 1005 & 218 & 0 \\ \hline
    {\bf Yield} & 0.39 & 0.50 & 0.11 & 0 \\ \hline
\end{tabular}
\end{center}

Table~2. {\small Relative yields of portraits belonging to anticorrelation (ACOR), market mill (MILL), correlation (COR) and anti-mill (AMILL) types
identified on the annual basis .}

\medskip

Let us stress that the relative yields of the $AC$ - classified patterns at NYSE and NASDAQ stock exchanges are markedly different, see Table 3

\medskip
\begin{center}
{\bf Table 3}
\end{center}
\begin{center}
\begin{tabular}{|c|c|c|}
  \hline
  {\bf Type} & {\bf NYSE} & {\bf NASDAQ} \\ \hline
  Anticorrelation & 0.15 & 0.64 \\ \hline
  Market Mill & 0.65 & 0.35 \\ \hline
  Correlation & 0.2 & 0.01 \\ \hline
  Anti-mill   & 0 & 0 \\ \hline
\end{tabular}
\end{center}

Table~3. {\small Yields of anticorrelation, market mill, correlation and anti-mill patterns at NYSE and NASDAQ stock exchanges}

\medskip

It is interesting to note that, as follows from Table~3, at NYSE the relative yields of different patterns are in agreement with the combinatorial
yields in Table~1, while at NASDAQ these proportions are heavily distorted.

\subsubsection{Subgroup portraits and conditional dynamics}

The examples discussed in the paragraph~\ref{visclas} correspond to "ideal" examples of market mill, correlation and anticorrelation behavior. Let us
now study the group portraits of market mill, anticorrelation and correlation subensembles formed according to the above-described $AC$
classification. These portraits are shown in Figs.~11-13\footnote{The plots are numbered in such a way that PNG files follow the EPS ones.}. Let us
also consider the shapes of the conditional mean response characterizing each of the groups under consideration. The result is shown in Fig.~7.
\begin{figure}[h]
 \begin{center}
 \epsfig{file=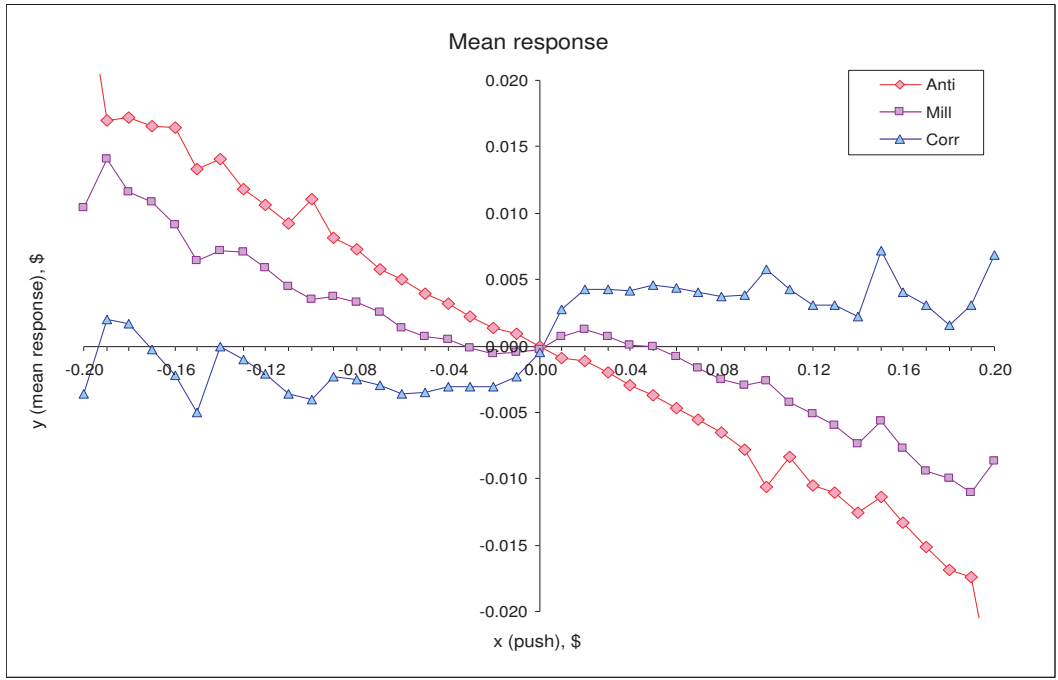,width=14cm}
 \end{center}
\caption{Mean response}
\end{figure}
We see that the conditional mean response for the mill group has a clear z-shaped structure. The anticorrelation group shows a monotonous dependence
on the push.  In this case the "extra" blades seen in sectors II and VI in Fig.~11 are too weak to change the mean behavior at small pushes. For the
correlation group the dependence of conditional mean response on the push is markedly nonlinear and can be described as stretched z-shaped one.

The basic conclusion one draws from analyzing the asymmetry patterns shown in Figs.~11-13 is that, at least within the framework of the $AC$ -
classification under consideration, pure correlation and anticorrelation patterns do not exist. They can rather be naturally described as "deformed"
market mill patterns, in which "wrong" sectors are never really empty. For anticorrelation group this means that a well - defined contrarian
conditional dynamics at large pushes coexists with a pronounced trend-following component that serves as an amplifier of the push. For the stocks in
correlation group the conditional dynamics is trend - following in the domain around the zero push, but at large pushes the behavior is again
contrarian. All three patterns under consideration are thus characterized by the specific mixture of correlative amplifying dynamics for small pushes
and anticorrelative at large ones.

\subsubsection{Pattern content on the monthly basis}

In the previous paragraph we have seen that for stocks classified as correlation and especially anticorrelation ones the group portraits differ from
the "ideal" ones. A possible reason for this is a temporal instability of the patterns. Let us thus consider  an evolution of the type of a stock on
a monthly basis. For the considered two-year period this generates, for each stock, a sequence of 24 symbols (a mixture of MILL,ACOR,COR). The type
of a stock can then be related to the dominant (highest frequency) pattern. For of 2000 stocks under consideration this procedure gives the following
results\footnote{For 51 stocks two highest frequencies were equal, so the type identification was not possible}

\begin{center}
\begin{tabular}{|c|c|c|c|c|}
\hline
    {\bf Type} & {\bf ACOR} & {\bf MILL} & {\bf COR} & {\bf AMILL} \\ \hline
    {\bf Number} & 836 & 969 & 144 & 0 \\ \hline
    {\bf Yield} & 0.41 & 0.48 & 0.07 & 0 \\ \hline
\end{tabular}
\end{center}

\medskip

{\bf Table 2} {\small Relative yields of portraits belonging to anticorrelation (COR), market mill (MILL), correlation (COR) and anti-mill (AMILL)
types computed on the monthly basis.}

The average proportion of time spent in the dominant configuration is $0.73$. It is interesting to note, that in the overwhelming majority of cases
the dominant pattern coexists with only one subdominant one.

\subsubsection{Indices}

Let us complete the analysis of this paragraph by considering the asymmetry properties of the bivariate distribution ${\cal P}(x,y)$ for two indices
and two corresponding ETFs : (SPX, NDX) and (SPY, QQQ) respectively. SPX and NDX are non-tradable, while SPY and QQQ are their tradable
counterparts.\footnote{The tradable indices are defined in such a way that SPY=SPX/40 and QQQ=NDX/10} The corresponding asymmetry patterns are shown
in Fig.~14. We find that
\begin{itemize}
 \item{Non-tradable indices NDX and SPX are characterized by correlation pattern (COR).}
 \item{Tradable indices QQQ and SPY are characterized by anticorrelation pattern (ACOR).}
\end{itemize}

\section{Pattern stability}

The effectiveness of the classification described in the previous section can naturally be measured by the stability of the individual portraits. The
visual stability of the asymmetry patterns can be very spectacular, see Fig.~17 in \cite{LTZZ05a}, in which we show the asymmetry patterns for three
sample stocks in two consequent semi-annual periods.

A development of a more quantitative description of pattern stability is possible at different levels of sofistication.

Crude quantitative characterization of the pattern stability can can be done in terms of the stability of the representative point in the $AC$ plane.
This is illustrated by plotting the trajectories of 10 sample stocks in the $AC$ plane in 2004, in which the year is subdivided into four subperiods,
in Fig.~8.
\begin{figure}[h]
 \begin{center}
 \leavevmode
 \epsfysize=12cm
 \epsfbox{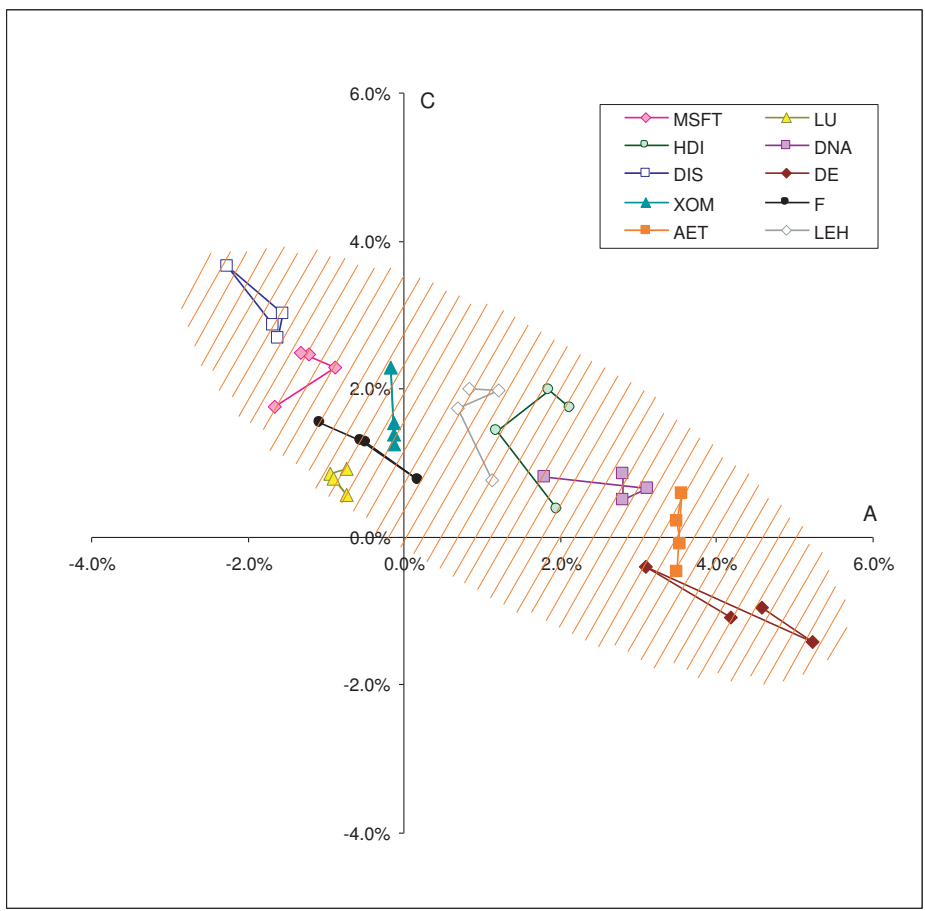}
 \end{center}
\caption{Trajectories of ten sample stocks in the $AC$ plane.}
\end{figure}
Fig.~8 illustrates a dominating tendency for the position of an $AC$ image of the stock pattern being remarkably stable. A shaded area in Fig.~8
shows a "domain of attraction" of the $AC$ - images in the $AC$ plane in which the predominant number of all 2000 images lies, cf. Fig.~6.

The above - introduced AC classification allows to monitor basic regime changes of a stock. A nontrivial example is provided by the quarterly
trajectory for the stocks having experienced an acquisition or merger in 2004-2005 shown in Fig.~9\footnote{We have considered the following
acquisitions:
\begin{itemize}
\item{Acquisition of The Gillette Company (G) by The Procter \& Gamble Co.(PG) announced on 01/28/2005.}
\item{Merger of The May Department Stores Company (MAY) with Federated Department Stores, Inc. (FD) announced on 02/28/2005.}
\item{Acquisition of Argosy Gaming Company (AGY) by Penn National Gaming, Inc. (PENN) announced on 11/03/2004.}
\end{itemize}
}.
\begin{figure}[h]
 \begin{center}
 \leavevmode
 \epsfysize=10cm
 \epsfbox{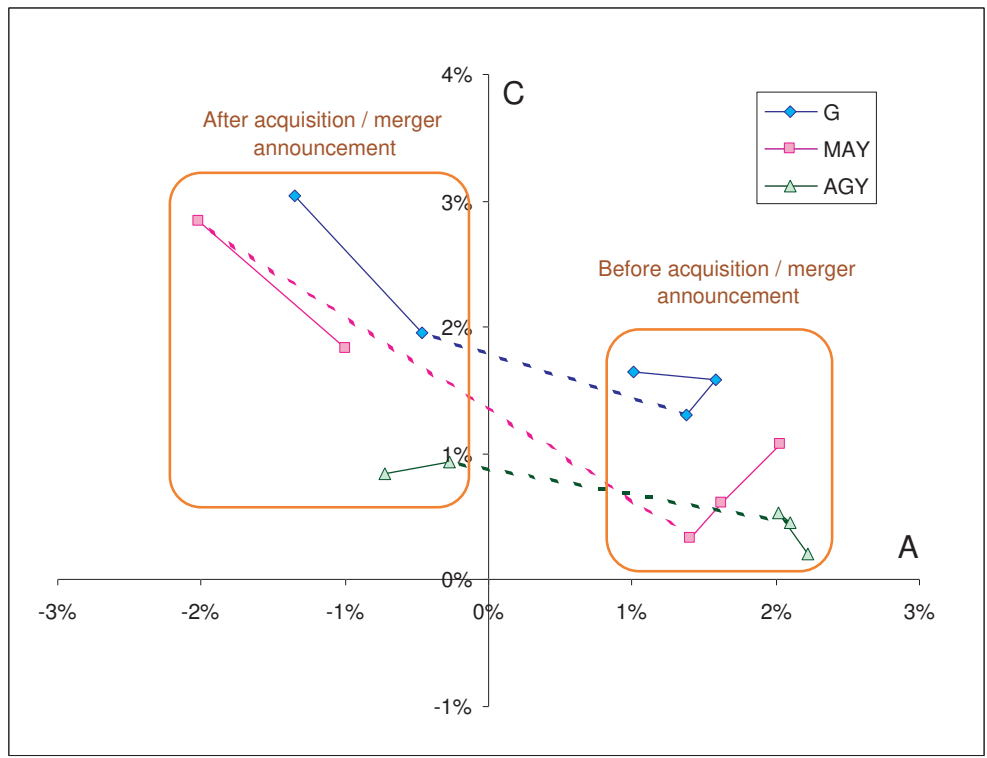}
 \end{center}
\caption{Trajectories of stocks having experienced an acquisition or merger in the $AC$ plane.}
\end{figure}
We see that for the three stocks considered (G, MAY and AGY) there is a clear change of pattern from being of market mill type before the merger
(acquisition) to the anticorrelation one after it.

A more detailed qualitative characterization of the stability of individual portraits in time can be done with the help of spider web diagrams.  A
stability of the pattern means that the spider web diagram for the same stock does not change much when computed for two nonintersecting time
periods. This is illustrated in Fig.~10, in which we show the stability of the spider's web patterns of DIS, HDI and DE (Fig.~10 (a) -- (c)) as well
as that of PBG, MAT and ACF patterns (Fig.~10 (d) -- (e)) chosen to illustrate that stability of spider's web pattern is a universal property not
restricted to conservation of "canonical" anticorrelation, market mill or correlation patterns. Especially pronounced is the stability of the highly
asymmetric and skewed pattern for PBG (Fig.~10 (d)).
\begin{figure}[h]
\begin{center}
\epsfig{file=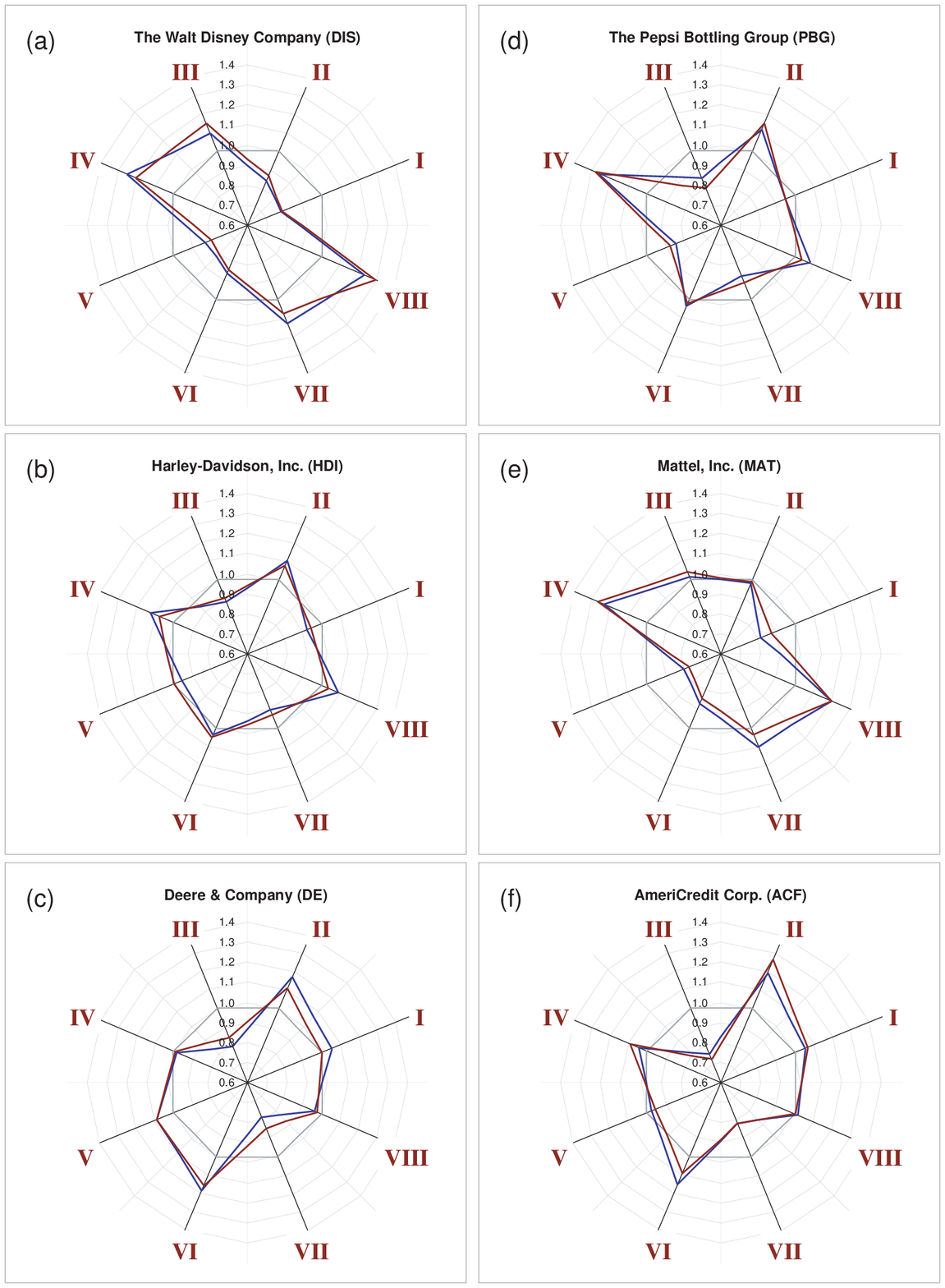,width=13.5cm}
\end{center}
\caption{Spider's web portraits for (a)~DIS; (b)~HDI; (c)~DE; (d)~PBG;  (e)~MAT; (f)~ACF. }
\end{figure}

To get a more quantitative estimate, let us introduce a distance in the pattern space as parametrized by the vectors of normalized sector weights
${\bold w} = c(w_1, \cdots , w_8)$, $d({\bf w}_1,{\bf w}_2) = {\rm Dist}_{L_1} ({\bf w}_1,{\bf w}_2)$. Each stock $i$ is characterized by the set
vectors ${\bf w}^{i}_1, \cdots {\bf w}^{i}_{N}$, where $N$ depends on the time scale at which one considers the pattern stability. A temporal
evolution of the pattern is then described by $N-1$ distances $d^i_{1\,2}, d^i_{2\,3}, \cdots , d^i_{N\,N-1}$,  where $d^i_{1\,2} \equiv d({\bf
w}^i_1,{\bf w}^i_2)$, etc. A stability of the individual pattern can be estimated by comparing the average distance  $\langle d_t^i \rangle = {\rm
mean} (d^i_{1\,2}, d^i_{2\,3}, \cdots , d^i_{N\,N-1})$ with an average {\it simultaneous} distance between the chosen pattern and the patterns of
other stocks in the same group  (mill, correlation or anticorrelation) $\langle d_{{\rm group}}^{i\,k} \rangle$ and between the pattern and all other
patterns $\langle d_{{\rm total}}^{i\,k} \rangle$, where in the latter two cases the averaging is first done over the simultaneously existing
patterns and then over time. The result can be compactly expressed through two ratios
\begin{equation}
\rho_G \, = \, \frac{\langle d_{{\rm group}}^{i\,k} \rangle}{ \left \langle \langle d_t^i \rangle \right \rangle } \,\,\,\,\ {\rm and} \,\,\,\,
\rho_T \, = \, \frac{\langle d_{{\rm total}}^{i\,k} \rangle}{ \left \langle \langle d_t^i \rangle \right \rangle }
\end{equation}
If one chooses the monthly time scale, one finds $\rho_G = 1.64 \pm 0.38$ and $\rho_T = 1.8 \pm 0.43$. This confirms that self-similarity of an
individual pattern is indeed a dominating feature of the data.

\section{Conclusions}

Let us formulate once again the main conclusions of the present paper.
\begin{itemize}
    \item{Visual inspection of bivariate dependence patterns allows to classify them into three major groups: correlation,
    anticorrelation and market mill.}
    \item{A suggested classification in terms of relative weights of different sectors in the push-response plane is shown to provide
    a stable characterization of the conditional dynamics of individual stocks. This stability is conveniently visualized by considering
    the corresponding spider web diagrams.}
    \item{A developed characterization of patterns in terms of a position in the $AC$ plane is shown to provide an adequate
    characterization of the type of conditional dynamics.}
    \item{Analysis of summary patterns for all of the three groups reveals common basic features of conditional dynamics:
    trend-following response at small push magnitudes and contrarian response at large push magnitudes.}
    \item{Quantitative classification of bivariate patterns reveals important differences between stocks traded at NYSE and
    NASDAQ stock exchanges.}
    \item{The asymmetry pattern characterizing non-tradable indices is of correlation type while that of tradable indices
    is of anticorrelation one.}
    \item{Specific pattern is shown to be a stable characteristics of the stock.}
\end{itemize}

\end{document}